# Analytical coupled vibroacoustic modeling of membrane-type acoustic metamaterials: plate model

**Yangyang Chen and Guoliang Huang** [a]

*Department of Systems Engineering, University of Arkansas at Little Rock, Little Rock, AR 72204, USA*

**Xiaoming Zhou and Gengkai Hu**

*Key Laboratory of Dynamics and Control of Flight Vehicle, Ministry of Education, School of Aerospace Engineering, Beijing Institute of Technology, Beijing, China, 100081*

**Chin-Teh Sun**

*School of Aeronautics and Astronautics, Purdue University, W. Lafayette, IN 47907, USA*

By considering the membrane's dissipation, the membrane-type acoustic metamaterial (MAM) has been demonstrated as a super absorber for low-frequency sound. In the paper, a theoretical vibroacoustic plate model is developed to reveal sound energy absorption mechanism within the MAM under a plane normal incidence. Based on the plate model in conjunction with the point matching method, the in-plane strain energy of the membrane due to the resonant and antiresonant motion of the attached masses can be accurately captured by solving the coupled vibroacoustic integrodifferential equation. Therefore, the sound absorption of the MAM is obtained and discussed, which is also in good agreement with the prediction from the finite element method. In particular, microstructure effects including eccentricity of the attached masses, the depth, thickness and loss factor of the membrane on sound absorption peak values are quantitatively investigated.

---

[a] Electronic mail: glhuang@ualr.edu

## I. INTRODUCTION

The attenuation/absorption of low frequency sound is of great interest for noise control. Common homogeneous materials, such as foam and composite panels, usually exhibit weak absorptions in the low frequency range, due to their dissipative power is quadratic in material velocities. Recently, membrane-type acoustic metamaterials (MAMs) have been suggested to possess excellent acoustic properties for sound insulation at 100-1000 Hz frequency regime, the most difficult regime as dictated by the mass density law.[1,2] This MAM comprises a pre-tensioned elastic rubber membrane attached with only one rigid circular mass. Nearly total reflection of low-frequency sound has been achieved.[1-5] To realize broadband wave attenuation and enhance the wave dissipation, it is usually necessary to increase the energy density inside the MAM through multiple resonators. Motivated by this idea, a thin elastic membranes decorated with designed patterns of multiple rigid platelets was further suggested.[6] The basic microstructure of this MAM consists of a membrane with multiple attached small heterogeneous masses acting as resonators with fixed boundaries imposed by a relatively rigid grid. It was reported that the one-layer MAM can absorb 86% of the acoustic waves at ~170 Hz, and absorb 99% with two layers at the lowest resonant frequency. However, the wave attenuation/absorption mechanism is not well interpreted and understood yet.

Issues about sound transmissions through membranes and partitions have been intensively investigated for decades.[7-10] A coupled vibroacoustic theoretical model of the sound transmission through a MAM has been proposed in the companion paper,[11] in which the classical membrane theory can precisely govern the motion of the pre-stressed thin elastic membrane. However, the dissipative/absorbed sound power, which is proportional to the total strain energy of the membrane, cannot properly be calculated by the classical membrane theory, because effects of the bending stiffness are neglected. Therefore, the flexural plate theory for the MAM will be highly needed for the purpose of the energy absorption calculation.

For vibrations of thin plates combining with varies boundary conditions, governing equations and the Galerkin procedure with several approximate series solutions have been suggested.[12-13] The problem of sound transmission through a thin plate based on vibroacoustic plate model has been solved with integrals of Green's functions.[14] Whereas, modeling vibrations and sound dissipations of the MAMs would address a challenging issue, in which a pre-stressed clamped thin plate carrying finite attached masses of arbitrary shapes needs to be solved. Galerkin procedure and Rayleigh-Ritz method are the most commonly used systematic approaches to study vibrations of plates with attached masses, in which the bending stiffness of the attached mass is usually ignored.[15-17] However, different from those studies, bending stiffness of attached masses on MAMs cannot be neglected. Instead, such attached masses would be rigid compared with the thin rubber membrane. A convenient and highly approximate point matching scheme has been employed to analyze free vibrations of MAMs, which assumed that interactions between masses and the membrane can be represented by distributed loadings on their boundaries.[11] Another issue about the MAM is geometric nonlinearities of the rubber membrane, in which in-plane pre-stresses are usually comparable with the Young's modulus. A plate theory considering incremental deformation and initial stress has been developed for orthotropic laminated plates.[18]

In this paper, to investigate sound absorptions of MAMs, we will develop a vibroacoustic plate model to accurately capture strain energy in the membrane. The initial tension induced effective bending stiffness is first derived by adopting nonlinear strains in the incremental energy method. Eigenfrequencies and eigenmodes of the MAM are then solved by using point matching scheme, where the Galerkin procedure with double consine series expansions is selected. Finally, the dissipative power is calculated through solving the coupled vibroacoustic integrodifferential equation with complex Young's modulus and the modal superposition method. Specifically, microstructure effects on sound absorptions are quantitatively investigated, which include eccentricities and numbers of masses, depth and thickness of the membrane and the membrane's loss factor.

## II. THEORETICAL PLATE MODEL

Consider now the unit cell of an MAM in a global Cartesian coordinate system ($x,y$) with the origin $O$ on the lower left corner of the rectangular membrane, as shown in Fig. 1 (a), where the membrane is symmetrically attached by several masses with respect to the central line of the membrane along the $x$ direction. Masses can be of arbitrary symmetric shapes with respect to the central line of the membrane along the *y* direction. The number of masses is denoted as *S*, and there are $I_s$ collocation points along edges of the *s*-th mass. In the figure, the membrane is subject to initial tension *T* per unit length uniformly in both *x* and *y* directions. The thickness, width, depth and density per unit area of the membrane are denoted as *h*, $L_x$, $L_y$ and $\rho_m$, respectively, and the weight of the *s*-th mass is $m_s$. An *s*-th local Cartesian coordinate system $(x'_s, y'_s)$ is assumed with origin $O'_s$ in the center of the gravity of the *s*-th mass. The *i*-th collocation point on the inner boundaries between the *s*-th mass and the membrane is denoted as $(X_i^{(s)}, Y_i^{(s)})$ in the global Cartesian coordinate system (*x*,*y*), and $(X_i^{(s)\prime}, Y_i^{(s)\prime})$ in the corresponding *s*-th local Cartesian coordinate system $(x'_s, y'_s)$. In the study, we focus on the sound absorption of the stretched MAM in a tube subject to a normally incident plane sound wave, as shown in Figure 1(b). Perfectly absorbing boundary conditions are assumed in both ends of the tube so that there will be no multiple reflected waves to the MAM.

## A. Effective bending stiffness of the plate with initial stress

We consider a flexural motion of a pre-tensioned elastic rubber membrane by using the thin plate model with uniform in-plane initial stress, $\sigma_0 = T/h$, in both *x* and *y* directions, and the amplitude of initial stresses are comparable with the Young's modulus of the rubber membrane. According to the Kirchhoff hypothesis, displacement fields in *x* and *y* directions can be expressed, respectively, as

$$u(x,y,z,t) = -z \frac{\partial w(x,y,t)}{\partial x}, \tag{1}$$

$$v(x,y,z,t) = -z \frac{\partial w(x,y,t)}{\partial y}, \tag{2}$$

where *z* denotes the coordinate measured from the neutral plane of the membrane, and *w* is the out-of-plane displacement of this neutral plane. Green-Lagrangian in-plane strains are considered and expressed by

$$\varepsilon_x = -z \frac{\partial^2 w}{\partial x^2} + \frac{1}{2}\left[\left(z\frac{\partial^2 w}{\partial x^2}\right)^2 + \left(z\frac{\partial^2 w}{\partial x \partial y}\right)^2 + \left(\frac{\partial w}{\partial x}\right)^2\right], \tag{3}$$

$$\varepsilon_y = -z \frac{\partial^2 w}{\partial y^2} + \frac{1}{2}\left[\left(z\frac{\partial^2 w}{\partial y^2}\right)^2 + \left(z\frac{\partial^2 w}{\partial x \partial y}\right)^2 + \left(\frac{\partial w}{\partial y}\right)^2\right], \tag{4}$$

$$\varepsilon_{xy} = -z\frac{\partial^2 w}{\partial x \partial y} + \frac{1}{2}\left[z^2 \frac{\partial^2 w}{\partial x^2}\frac{\partial^2 w}{\partial x \partial y} + z^2 \frac{\partial^2 w}{\partial y^2}\frac{\partial^2 w}{\partial x \partial y} + \frac{\partial w}{\partial x}\frac{\partial w}{\partial y}\right]. \tag{5}$$

The out-of-plane shear strains and stresses are neglected. Due to the free surfaces of the plate, we have $\sigma_z = 0$. If we assume the incremental deformation is infinitesimal, the final state of stress can be given by the Trefftz stress components as

$$\sigma_x = \sigma_0 + \frac{E}{(1-v^2)}(\varepsilon_x + v\varepsilon_y), \tag{6}$$

$$\sigma_y = \sigma_0 + \frac{E}{(1-v^2)}(\varepsilon_y + v\varepsilon_x), \tag{7}$$

$$\sigma_{xy} = G\varepsilon_{xy}, \tag{8}$$

in which $G = E/2(1 + v)$ with $E$ and $v$ being the Young's modulus and Poisson's ratio of the stretched elastic membrane.

Based on the linear constitutive relation in Eqs. (6-8), the strain energy per unit initial volume is

$$PE = \sigma_0(\varepsilon_x + \varepsilon_y) + \frac{E}{2(1-v^2)}\left[\varepsilon_x^2 + \varepsilon_y^2 + 2v\varepsilon_x\varepsilon_y + 2(1-v)\varepsilon_{xy}^2\right]. \tag{9}$$

The incremental strain energy is

$$\Delta PE = PE - \sigma_0(\bar{\varepsilon}_x + \bar{\varepsilon}_y), \tag{10}$$

where $\bar{\varepsilon}_x = \partial u/\partial x$ and $\bar{\varepsilon}_y = \partial v/\partial y$ are usual linear strains.[19]

Substituting Eqs. (3-5) and Eq. (9) into Eq. (10), the total incremental strain energy per unit initial area can be obtained by integrating $\Delta PE$ over the plate thickness and neglecting terms with three or higher order powers in displacement-gradients as

$$U = \frac{1}{2}\left\{\frac{\sigma_0 h^3}{12}\left[\left(\frac{\partial^2 w}{\partial x^2}\right)^2 + \left(\frac{\partial^2 w}{\partial y^2}\right)^2 + 2\left(\frac{\partial^2 w}{\partial x \partial y}\right)^2\right] + T\left[\left(\frac{\partial w}{\partial x}\right)^2 + \left(\frac{\partial w}{\partial y}\right)^2\right] + D\left\{(\nabla^2 w)^2 - 2(1-v)\left[\frac{\partial^2 w}{\partial x^2}\frac{\partial^2 w}{\partial y^2} - \left(\frac{\partial^2 w}{\partial x \partial y}\right)^2\right]\right\}\right\}, \tag{11}$$

in which $D = \frac{Eh^3}{12(1-v^2)}$. The kinetic energy per unit initial area of the membrane is expressed as

$$K = \frac{1}{2}\rho_m \left(\frac{\partial w}{\partial t}\right)^2. \tag{12}$$

By applying Hamilton's principle on Eq. (11) and Eq. (12), the governing equation of flexural motion of pre-stressed plate can be written as

$$D^*\nabla^4 w - T\nabla^2 w + \rho_m \frac{\partial^2 w}{\partial t^2} = 0, \tag{13}$$

where $\nabla^4 = \nabla^2(\nabla^2)$, $\nabla^2 = \frac{\partial^2}{\partial x^2} + \frac{\partial^2}{\partial y^2}$, and $D^* = D + \frac{\sigma_0 h^3}{12}$, which is the effective bending stiffness of the plate.

## B. Eigenvalue problem of the MAM

The attached masses are assumed to be rigid and perfectly bonded to the rectangular membrane. To properly capture effects of those masses on the deformation of the membrane, the point matching scheme is applied such that each mass can be represented by several point loadings on the membrane along their interfacial boundaries. For free vibration of the MAM, the governing equation of the rectangular membrane can be written as

$$D^*\nabla^4 w(x,y,t) - T\nabla^2 w(x,y,t) + \rho_m \frac{\partial^2 w(x,y,t)}{\partial^2 t} = \sum_{s=1}^{S}\sum_{i=1}^{I_s} F_i^{(s)}(t)\delta\left(x - X_i^{(s)}\right)\delta\left(y - Y_i^{(s)}\right), \tag{14}$$

where $F_i^{(s)}$ in the right hand side is the force loading at the $i$-th collocation points from the $s$-th attached mass along $s$-th inner boundaries, and $\delta$ is the Dirac delta function. Since only the steady-state response field will be considered, the time factor $e^{i\omega t}$, which applies to all the field variables, will be suppressed in the paper. Then, $F_i^{(s)}$ becomes a constant to be determined. For a clamped plate, the boundary conditions are

$$w = \frac{\partial w}{\partial x} = 0, \text{ on } x = 0, x = L_x, \tag{15}$$

$$w = \frac{\partial w}{\partial y} = 0, \text{ on } y = 0, y = L_y. \tag{16}$$

To solve Eq. (14), the Galerkin procedure is applied to seek an approximate solution. Since a plane sound wave can only induce symmetric modes on the MAM, we choose the double cosine series expansion to satisfy boundary conditions in Eq. (15) and Eq. (16) as

$$w = \sum_{m=1}^{\infty} \sum_{n=1}^{\infty} W_{mn} q_{mn}, \tag{17}$$

$$W_{mn} = \left(1 - \cos\frac{2m\pi x}{L_x}\right)\left(1 - \cos\frac{2n\pi y}{L_y}\right). \tag{18}$$

Substituting Eq. (17) into Eq. (14), multiplying each term by $W_{mn}$, and integrating all terms over the domain ($0 \leq x \leq L_x, 0 \leq y \leq L_y$), lead to a linear system of equations for $q_{mn}$. Solutions of $q_{mn}$ can be expressed by the summation of $F_i^{(s)} q_{mni}^{(s)}$ from all the point loadings, where $q_{mni}^{(s)}$ is the solution of the linear system of equations with $F_i^{(s)} \delta\left(x - X_i^{(s)}\right)\delta\left(y - Y_i^{(s)}\right)$ in the right hand side. Then, the solution to Eq. (14) can be assumed as

$$w = \sum_{s=1}^{S} \sum_{i=1}^{I} F_i^{(s)} Q_i^{(s)}, \tag{19}$$

where $Q_i^{(s)} = \sum_{m=1}^{\infty} \sum_{n=1}^{\infty} q_{mni}^{(s)} \left(1 - \cos\frac{2m\pi x}{L_x}\right)\left(1 - \cos\frac{2n\pi y}{L_y}\right)$.

The unknown loading $F_i^{(s)}$ can be determined through the inner boundary conditions of the membrane. For the *s*-th attached mass with finite dimension, the general symmetric motion, which is composed of the rigid translation and rotation, can be expressed as

$$w_s'(x_s') = \bar{a}^{(s)} x_s' + \bar{c}^{(s)}, \tag{20}$$

in which $\bar{a}^{(s)}$ and $\bar{c}^{(s)}$ are two unknown constants. Equations of the motion of the *s*-th mass can be written as

$$-\sum_{i=1}^{I_s} F_i^{(s)} = m_s \left.\frac{\partial^2 w_s'}{\partial t^2}\right|_{(x_s'=0)} = -m_s \omega^2 \bar{c}^{(s)}, \tag{21}$$

$$-\sum_{i=1}^{I_s} F_i^{(s)} X_i^{(s)\prime} = I_{y_s'}^{(s)} \frac{\partial^2 \psi_{y_s'}^{\prime(s)}}{\partial t^2} = -I_{y_s'}^{(s)} \omega^2 \bar{a}^{(s)}, \tag{22}$$

where $\psi_{y_s'}^{\prime(s)}$ and $I_{y_s'}^{(s)}$ are the rotational displacement with respect to the local $y_s'$ axis and the moment of inertia with respect to the local $y_s'$ axis of the *s*-th mass, respectively. From Eq. (21) and Eq. (22), two unknown constants $\bar{a}^{(s)}$ and $\bar{c}^{(s)}$ can be determined in terms of $F_i^{(s)}$.

Finally, based on displacement continuity on the collocation point, we have

$$w_s'\left(X_j^{(s)\prime}, Y_j^{(s)\prime}\right) - w\left(X_j^{(s)}, Y_j^{(s)}\right) = 0. \tag{23}$$

Substituting Eqs. (19) and (20) into Eq. (23) yields

$$\sum_{s=1}^{S}\sum_{i=1}^{I} F_i^{(s)} \left(\bar{a}_i^{(s)} X_j^{(s)\prime} + \bar{c}_i^{(s)} - Q_i^{(s)}\left(X_j^{(s)}, Y_j^{(s)}\right)\right) = 0. \tag{24}$$

where $\bar{c}_i^{(s)} = \frac{1}{m_s \omega^2}$ and $\bar{a}_i^{(s)} = \frac{X_i^{(s)\prime}}{I_{y_s'}^{(s)} \omega^2}$.

By applying Eq. (24) on all the collocation points and setting the determinant of the coefficient matrix in the linear system of equations for $\{F_i^{(s)}\}$ of size $(I \cdot S \times I \cdot S)$ to zero, natural frequencies of the MAM can be obtained. The corresponding mode shapes' weak form solutions can then be determined by solving a homogenous system of equations.

## C. Vibroacoustic modeling of the MAM

Consider a plane sound wave is normally incident on the MAM. The objective is to determine the dissipated power within the MAM. The governing equation of the acoustic excited membrane based on the plate theory can be expressed as

$$D^* \nabla^4 w - T \nabla^2 w - \omega^2 \rho_m w = p_1|_{(z=0)} - p_2|_{(z=0)} + \sum_{s=1}^{S}\sum_{i=1}^{I_s} F_i^{(s)} \delta\left(x - X_i^{(s)}\right)\delta\left(y - Y_i^{(s)}\right), \tag{25}$$

where $p_1$ and $p_2$ are pressures on the left and right surfaces of the MAM.

By combining equations in the acoustic field as stated in the paper,[11] Eq. (25) can be rewritten as

$$D^* \nabla^4 w - T \nabla^2 w - \omega^2 \rho_m w + 2i\omega \rho_1 c_1 \langle w \rangle - 2\omega^2 \rho_1 \int_0^{L_x} \int_0^{L_y} G(w - \langle w \rangle) dx^* dy^* = 2P_I +$$
$$\sum_{s=1}^{S}\sum_{i=1}^{I_s} F_i^{(s)} \delta\left(x - X_i^{(s)}\right)\delta\left(y - Y_i^{(s)}\right), \tag{26}$$

in which the Green function $G = \frac{e^{ik_1 S}}{4\pi S} + \frac{e^{ik_1 S_1}}{4\pi S_1}$ with $S = \sqrt{(x-x^*)^2 + (y-y^*)^2 + (z-z^*)^2}$, $S_1 = \sqrt{(x-x^*)^2 + (y-y^*)^2 + (z+z^*)^2}$, $\langle \cdot \rangle$ denotes the average of the parameter, $\rho_1$, $c_1$ and $k_1$ are the density, sound speed and wave number of air, respectively.

To solve the integrodifferential equation, we use the modal superposition method (Galerkin procedure) such that the motion of the membrane is assumed to be

$$w(x,y) = \sum_{k=1}^{+\infty} W_k \hat{q}_k, \tag{27}$$

where $\hat{q}_k$ is the unknown constant to be determined, and $W_k$ is the $k$-th order mode shape from the weak form solutions.

By substituting Eq. (27) into Eq. (26), multiplying each term with $W_l$ and integrating over the whole area of the rectangular membrane, combining Eq. (21) and Eq. (22) and the weak form solution $W_k$, and considering the orthogonality of eigenfunctions, we have

$$(\omega_l^2 - \omega^2)(\rho_s L_x L_y \langle W_l^2 \rangle)\hat{q}_l + i\text{Im}(D^*)\sum_{k=1}^{+\infty}\int_0^{L_x}\int_0^{L_y} \nabla^4 w_k \cdot w_l dxdy \cdot \hat{q}_k +$$

$$\sum_{k=1}^{+\infty} 2i\omega\rho_1 c_1 L_x L_y \langle W_l\rangle\langle W_k\rangle \hat{q}_k - \sum_{k=1}^{+\infty} 2\omega^2 \rho_1 \int_0^{L_x}\int_0^{L_y} W_l \int_0^{L_x}\int_0^{L_y} G(W_k - \langle W_k\rangle)dx^*dy^* dxdy \cdot$$

$$\hat{q}_k = 2P_I L_x L_y \langle W_l\rangle, \tag{28}$$

in which $\rho_s$ is the mass density per unit area of the MAM. The unknown constant $\hat{q}_k$ can be solved from the above linear system of equations. Then, the displacement field of the MAM can be obtained.

Recalling equations of 0-th order acoustic fields in the companion paper,[11] definitions and relation between transmission and reflection coefficients are

$$\tilde{T} = \frac{P_T}{P_I} = \frac{i\omega\rho_1 c_1 \langle W\rangle}{P_I}, \tag{29}$$

$$\tilde{R} = \frac{P_R}{P_I}, \tag{30}$$

$$\tilde{R} = 1 - \tilde{T}, \tag{31}$$

with $P_I$, $P_R$ and $P_T$ being complex amplitudes of incident, reflected and transmitted plane waves. The intensity transmission and reflection coefficients are

$$T_I = |\tilde{T}|^2, \tag{32}$$

$$R_I = |\tilde{R}|^2, \tag{33}$$

from which the dissipated power of the MAM can be calculated as

$$A_I = 1 - T_I - R_I = 2\left(\text{Re}(\tilde{T}) - \text{Re}(\tilde{T})^2 - \text{Im}(\tilde{T})^2\right). \tag{34}$$

where $\left|\text{Re}(\tilde{T})\right| \leq 1$ and $\left|\text{Im}(\tilde{T})\right| \leq 1$ according to the definition of acoustic transmission and reflection coefficients. Therefore, it can be easily derived that the maximum dissipated power $A_I$ cannot be greater than 50% for any thin MAM.

## III. VALIDATION OF THE THEORETICAL MODELING

To verify the developed vibroacoustic plate model, acoustic and vibration properties of the MAM from the current model are compared with those from the commercial finite element software, COMSOL Multiphysics, in which the acoustic-solid interaction with geometric nonlinearities is selected. The MAM consists of a membrane symmetrically attached with two semicircular platelets, as shown in Fig. 2. Clamped boundary conditions are applied on all edges of the MAM, and the rigid wall boundary condition is used for the side boundary of the air. Two acoustic radiation boundaries are assumed on both ends of the system. Material properties and geometrical dimensions of the membrane and attached masses are given in Table 1. The loss factor of the rubber is set to be $\chi_0 \omega$ with $\chi_0 = 4.2 \times 10^{-4}$ s. For properties of air, $\rho_1 = 1.29$ kg/m$^3$ and $c_1 = 340$ m/s. The convergence of the finite element analysis is firstly conducted through analysis of absorption coefficients and displacement amplitudes at the first absorption peak frequency with different meshes, as shown in Figs. 3(a) and (b), respectively. It can be found that the numerical results are convergent when the number of total degree of freedoms (DOFs) of the system approaches 600 thousands.

Fig. 4 shows the comparison of intensity transmission, reflection and absorption coefficients of the MAM from both the theoretical model and the finite element analysis. In the theoretical model, the number of collocation points for one half of the semicircular mass is set as $I = 20$, and the number of cosine series expansions are truncated as $M = N = 40$ to make the result convergent. It can be seen that our analytical results (solid line) agree well with those from the finite element method (dash line). Three absorption peaks, located in 190 Hz, 356 Hz and 727 Hz from the analytical model and 190 Hz, 344 Hz and 710 Hz from the finite element method, together with three transmission peaks are observed near resonant frequencies, which are 189 Hz, 356 Hz and 733 Hz from the analytical model and 191 Hz, 337 Hz and 738 Hz from the finite element method. The dissipated power (absorption coefficient) at the three absorption peaks is calculated to be 37, 31 and 26% from the theoretical model and 41, 22 and 29% from the finite element method.

To validate the capacity of the current model for the energy absorption, the displacement amplitude and strain energy density within the mid-plane of the membrane at three absorption peak frequencies predicted from the current model and the finite element method are compared at Fig. 5 and Fig. 6. The images in Fig. 5 (a) and Fig. 6 (a) are from the analytical model, and the images in Fig. 5 (b) and Fig. 6 (b) are from the finite element method. The strain energy density within the mid-plane of the membrane in the 2-D theoretical model is calculated by averaging the strain energy density within the plate through the thickness. The color bar in the figure represents a logarithmic scale for the strain energy density. Good agreement between the analytical and numerical results is observed in both Fig. 5 and Fig. 6. As shown in Fig. 5, for the MAM with two attached semicircular masses, the first absorption peak is caused by both the translational and rotational motion of the masses, whereas the second peak is mainly caused by the rotational motion of the masses. The third peak is caused by the strong vibration of the partial membrane between the two masses. From Fig. 6, it can be seen that the strain energy density in the perimeter and clamped regions of the membrane is extremely much higher than the other regions by about three orders of magnitude at all the three absorption peaks. According to the fact that the local dissipated power is proportional to the strain energy density, most of the absorbed sound energy would be dissipated in these regions. The largest absolute discrepancy (around 10%) of the absorption at the second peak can be attributed to the approximation of the Kirchhoff hypothesis, which assumes that in-plane shear stains are dependent on out-of-plane displacement. Overall, it is clearly evident that the proposed model can accurately capture the sound energy dissipation behavior of the MAM as those in the finite element method.

The loss factor is usually determined by fitting theoretical absorptions with experimental absorptions. Effects of loss factors will be discussed in the next section. However, as illustrated in Eq. (33), dissipations would not be greater than 50% with any loss factors. Resulted theoretical absorptions cannot be as high as those in the experimental results as shown in Ref. 6. These higher experimental absorptions are mainly caused by the possible out-of-plane curvature in the frame, which support and clamp the MAM, and the imperfect symmetry of attached masses. The curvature of the frame can definitely lead to an oblique incidence, which may enlarge absorptions. The imperfect symmetry can contribute to absorptions through some exited asymmetric modes, which are not properly accounted in both analytical model and the finite element method.

## IV. RESULTS AND DISCUSSIONS

Based on the developed analytical model, we will investigate effects of the eccentricity of masses, the width and thickness of the membrane and loss factors on the sound absorption behavior of the above MAM. The MAM attached with multiple semicircular masses will be also considered. The purpose of this study is to develop an accurate and highly effective analytical tool to optimize the design of MAMs on sound dissipations.

### A. The MAM with two semicircular masses

In practice, eccentricity of attached masses is a critical parameter that can be easily changed to fulfill design requirement of an MAM. Figure 7 shows effects of eccentricities of two symmetric semicircular masses on sound absorptions of the MAM. In the figure, the material and geometric properties of the MAM are the same as listed in Table 1 with $\chi_0 = 4.2 \times 10^{-4}$ s, and only the eccentricity of attached masses is changed. It can be found that the first absorption peak value is increased with the increase of the eccentricity. However, the third absorption peak value is decreased with the increase of the eccentricity. It is understandable that when the eccentricity is increased, the membrane curvature around circular edges of attached masses and two vertically clamped edges will become larger at the first resonant frequency, where masses vibrate strongly with both translational and rotational motion. The strain energy density in these regions will then become higher, therefore, the total absorption will increase at the first resonance frequency. The decrease of the third peak is caused by the reduced membrane curvature along straight edges of masses, in which the highest strain energy density concentrates by a strong vibration of the membrane. The second peak is increased slightly, when $d$ is changed from 6.5mm to 7.5mm, and is reduced from 31% to 18%, when $d$ is changed from 7.5mm to 8.5mm. The sharp drop of the second peak is due to the decrease of rotational displacement amplitudes of attached masses and membrane curvatures in parameter regions. It should be mentioned that the eccentricity can also affect the resonant frequencies of the MAM.

The membrane's width effects on sound absorption of the MAM are illustrated in Fig. 8. In the figure, the material and geometric properties of the MAM are the same as listed in Table 1 with $\chi_0 = 4.2 \times 10^{-4}$ s, and only the width of the membrane is changed. As shown in the figure, when the width is reduced from 16 mm to 14 mm, the first absorption peak is raised from 28% to 47%, and the third peak is increased from 25% to 29%, whereas the second peak is decreased from 34% to 27%. All of the three resonant frequencies will also be increased slightly. The increased absorption values at the first and third resonant frequencies are caused by the increased curvature in sharp corners of masses. The decreased absorption is due to the reduced rotational displacement amplitudes of attached masses at the second resonant frequency.

Figure 9 shows effects of membrane's thickness on sound absorption of the MAM. In the figure, the material and geometric properties of the AM are the same as listed in Table 1 with $\chi_0 = 4.2 \times 10^{-4}$ s, only the thickness of the membrane is changed. It can be seen that the first and third sound absorption peaks are increased when the membrane becomes thicker. The second sound absorption peak, however, is reduced. The increased absorptions can be attributed to the increase of the bending stiffness of the plate, which is proportional to the strain energy density of the plate. Nevertheless, the thicker membrane would confine rotational motion of attached masses, and eventually the sound absorption of the MAM will be reduced at the second resonant frequency.

Figure 10 illustrates effects of $\chi_0$, a constant of the loss factor, on sound absorption of the MAM. In the figure, material and geometric properties of the MAM are the same as listed in Table 1, only $\chi_0$ is changed. It can be easily observed that the 50% limit of sound absorption of the MAM is further verified numerically. As expected, absorption at most of the frequency range can be increased by raising the value of $\chi_0$, until the absorption reaches its limit. It is understandable that the larger loss factor usually means more energy can be damped and absorbed within the membrane. However, when the dissipated power has reached the limit with the increase of $\chi_0$, it cannot be increased anymore, and will be decreased instead with the increase of $\chi_0$.

## B. The MAM with four semicircular masses

Sound absorption of the MAM attached with four semicircular masses, as shown in Fig. 11, are investigated by the developed vibroacoustic plate model. In the figure, the eccentricities of inner and outer masses, $d_1$ and $d_2$ and the length of the membrane are selected as 7.5 mm, 16.5 mm and 49.0 mm, respectively. Other material and geometric properties of the MAM are the same as listed in Table 1. Sound absorptions of the MAM are plotted in Fig. 12 in function of $\chi_0$. When four attached masses are used, two additional absorption peaks are obviously observed for $\chi_0 = 2.1 \times 10^{-4}$ s. It should be mentioned that the second peak becomes less distinct when $\chi_0$ is above 4.2E-4 s. Effects of $\chi_0$ on the other peaks are similar as those of the MAM with two attached masses. The displacement amplitude fields in the MAM at the five peak frequencies are shown in Fig. 13. It can be found that the first four absorption peaks of the MAM are caused by both translational and rotational motions of the attached masses, whereby the fifth absorption peak is caused by a strong vibration of the membrane. It can be concluded that the more masses the MAM attached, the more resonant frequencies can be found in the low frequency range, which can produce more sound absorption peaks and make the spectrum of sound absorptions broader eventually.

## V. CONCLUSIONS

The vibroacoustic plate model is first developed to study sound absorptions and energy dissipations within MAMs under a normal incidence. The incremental energy method is applied to derive the effective bending stiffness of plates with initial in-plane stresses. Based on the plate model in conjunction with the point matching method, the in-plane strain energy of the membrane due to the resonant and antiresonant motion of the attached masses can be accurately captured by solving the coupled vibroacoustic integrodifferential equation. Therefore, the sound absorption of the MAM is obtained and discussed. The accuracy of the model is verified by comparison with the finite element method. Finally, parameter studies including masses' eccentricities and the width, thickness and the loss factor of the membrane on the sound absorption behaviors of the MAM with multiple attached masses are initially demonstrated.

**ACKNOWLEDGMENTS**

The authors would like to thank Dr. Ping Sheng from Hong Kong University of Science and Technology and Dr. Jun Mei from South China University of Technology, whose comments and discussions improved this work significantly. This work was supported in part by the Air Force Office of Scientific Research under Grant No. AF 9550-10-0061 with Program Manager Dr. Byung-Lip (Les) Lee and by National Natural Science Foundation of China under Grants 11221202, 11290153 and 11172038.

**Tables**

Tab. 1 Material properties and geometric parameters (2 masses)

|  | Membrane | Mass |
|---|---|---|
| Mass Density (kg/m$^3$) | 980 | 7870 |
| Young's modulus (Pa) | 1.9×10$^6$ | 2×10$^{11}$ |
| Poisson ratio | 0.48 | 0.30 |
| Thickness (mm) | 0.2 | 1 |
| Width (mm) | 31 | - |
| Height (mm) | 15 | - |
| Radius (mm) | - | 6 |
| Eccentricity $d$ (mm) | - | 7.5 |
| Pretension (N/m) | 44.0 | 0 |

*Figure Captions:*

FIG. 1. (a) MAM symmetrically attached with multiple masses of arbitrary symmetric shapes. (b) MAM subjected to a normal acoustic loading in a tube.

FIG. 2. MAM symmetrically attached with two semicircular masses.

FIG. 3. (a) Convergence analysis of the finite element method: Absorption coefficient. (b) Convergence analysis of the finite element method: Displacement amplitudes at the first resonant frequency.

FIG. 4. Validation of transmissions, reflections and absorptions of the MAM with two semicircular masses.

FIG. 5. Validation of displacement amplitude at absorption peak frequencies of the MAM with two semicircular masses.

FIG. 6. Validation of strain energy density at absorption peak frequencies of the MAM with two semicircular masses.

FIG. 7. Effects of masses' eccentricities to sound absorptions of the MAM with two semicircular masses.

FIG. 8. Effects of membrane's width to sound absorptions of the MAM with two semicircular masses.

FIG. 9. Effects of membrane's thickness to sound absorptions of the MAM with two semicircular masses.

FIG. 10. Effects of membrane's loss factors to sound absorptions of the MAM with two semicircular masses.

FIG. 11. MAM symmetrically attached with four semicircular masses.

FIG. 12. Sound absorptions of the MAM with four semicircular masses with different loss factors.

FIG. 13. Displacement amplitude at absorption peak frequencies of the MAM with four semicircular masses.

**Figures**

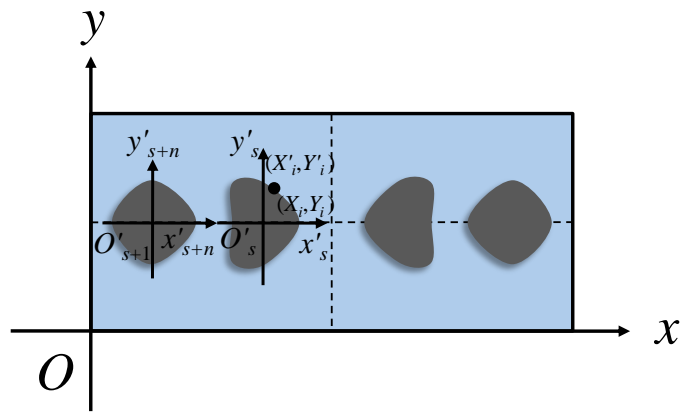

FIG. 1 (a)

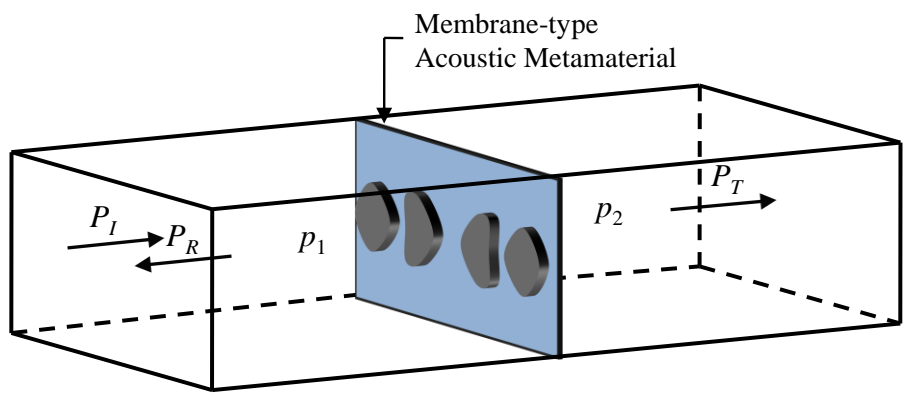

FIG. 1 (b)

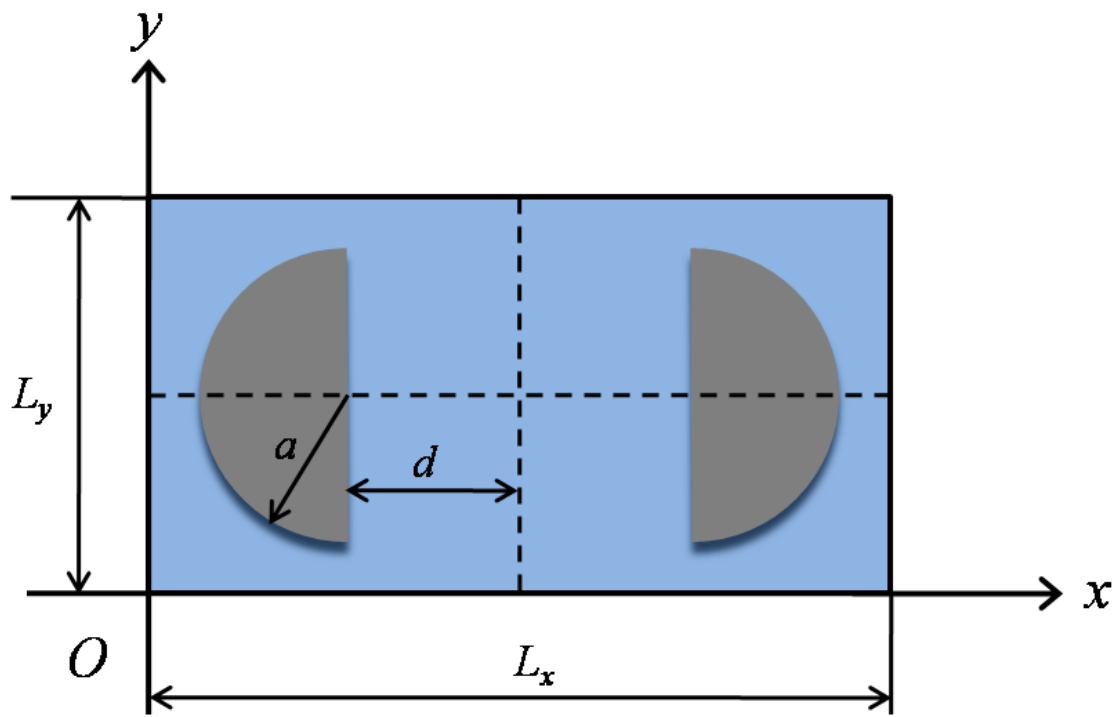

FIG. 2

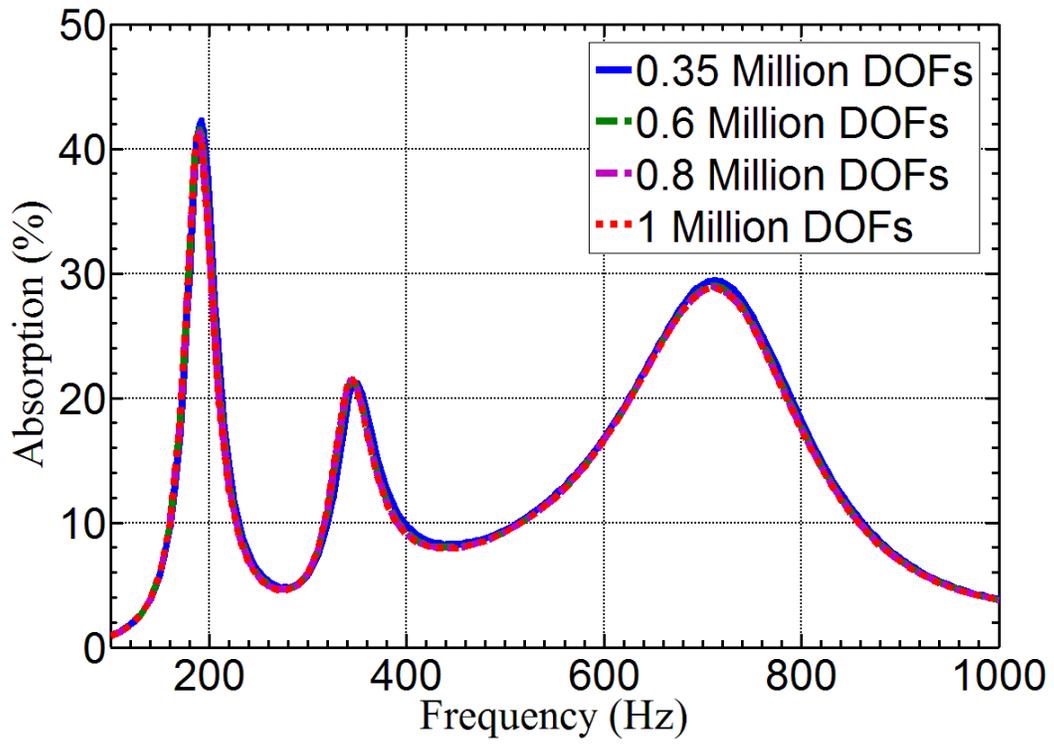

FIG. 3 (a)

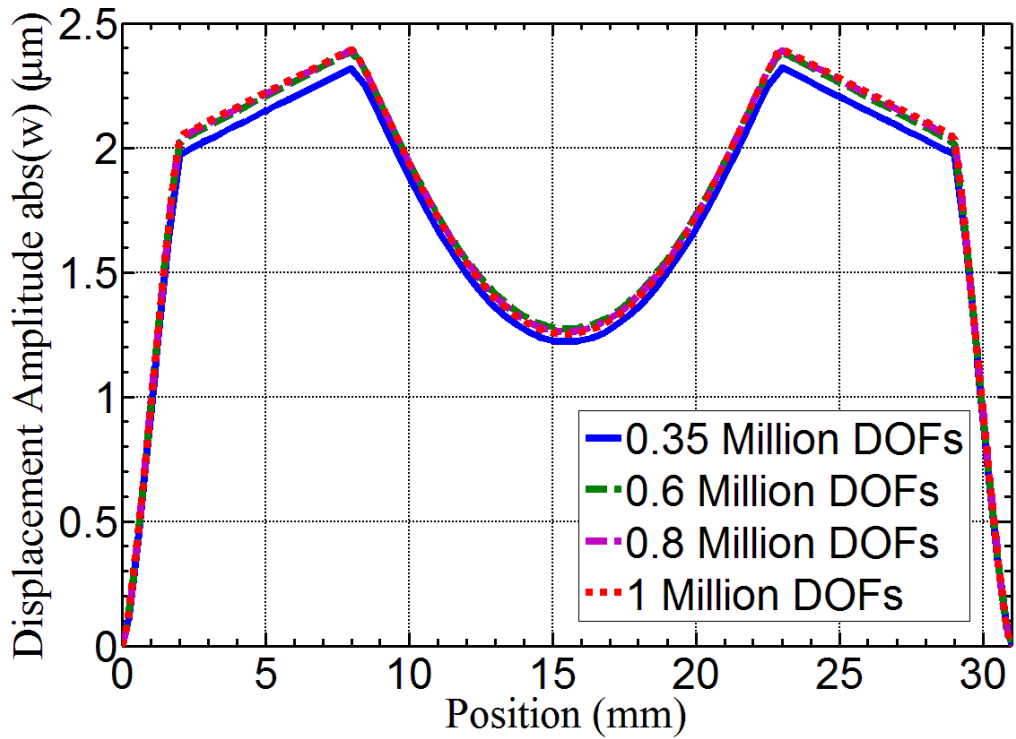

FIG. 3 (b)

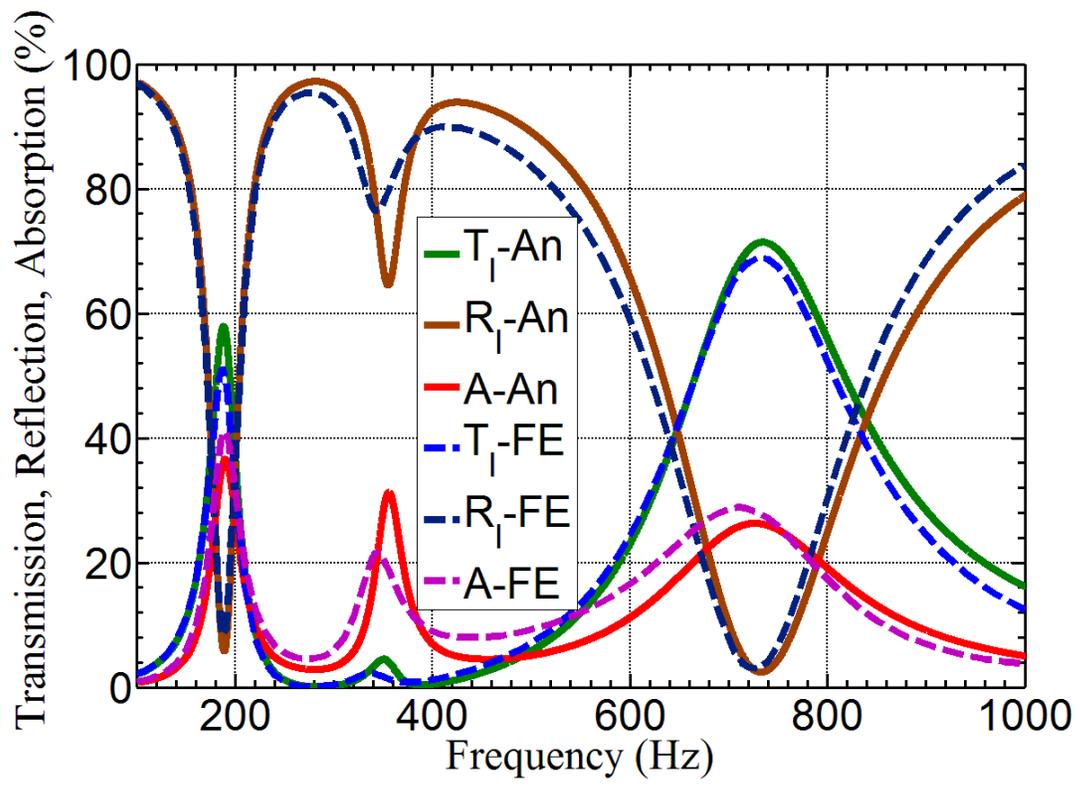

FIG. 4

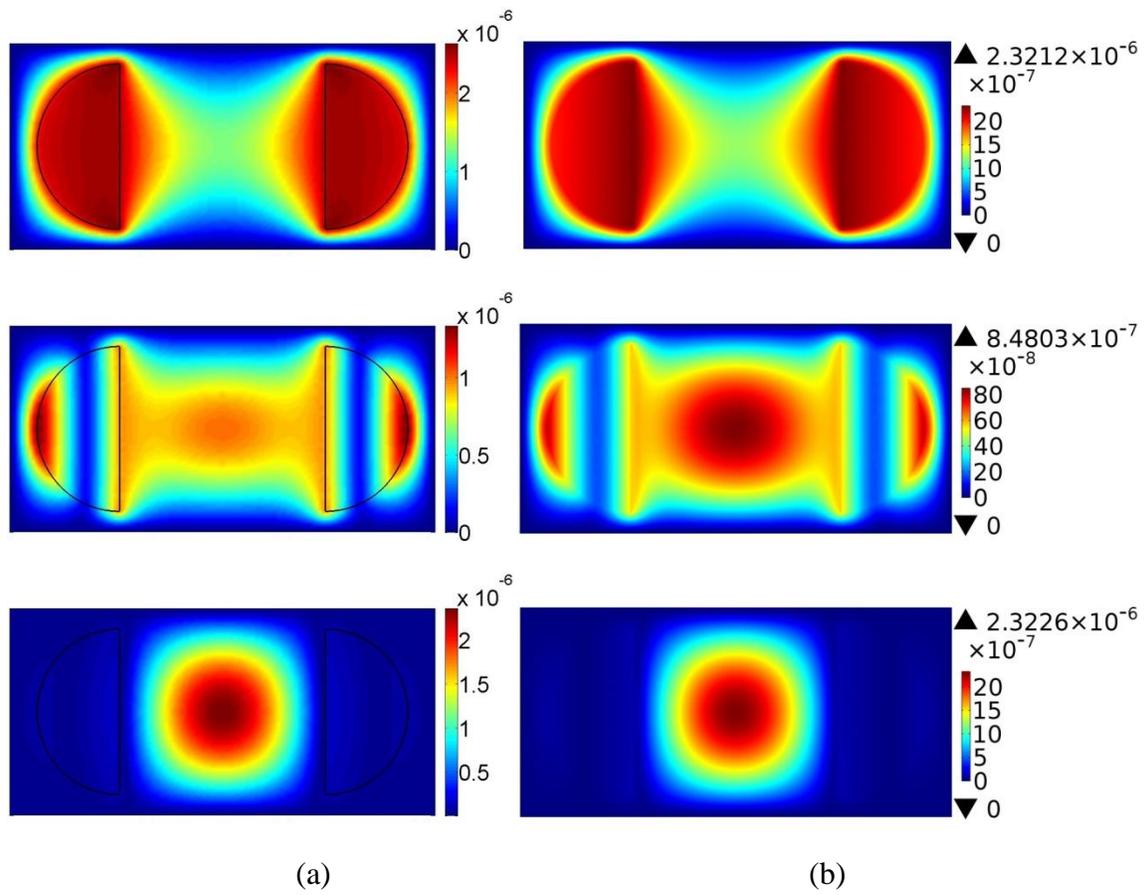

(a)            (b)

FIG. 5

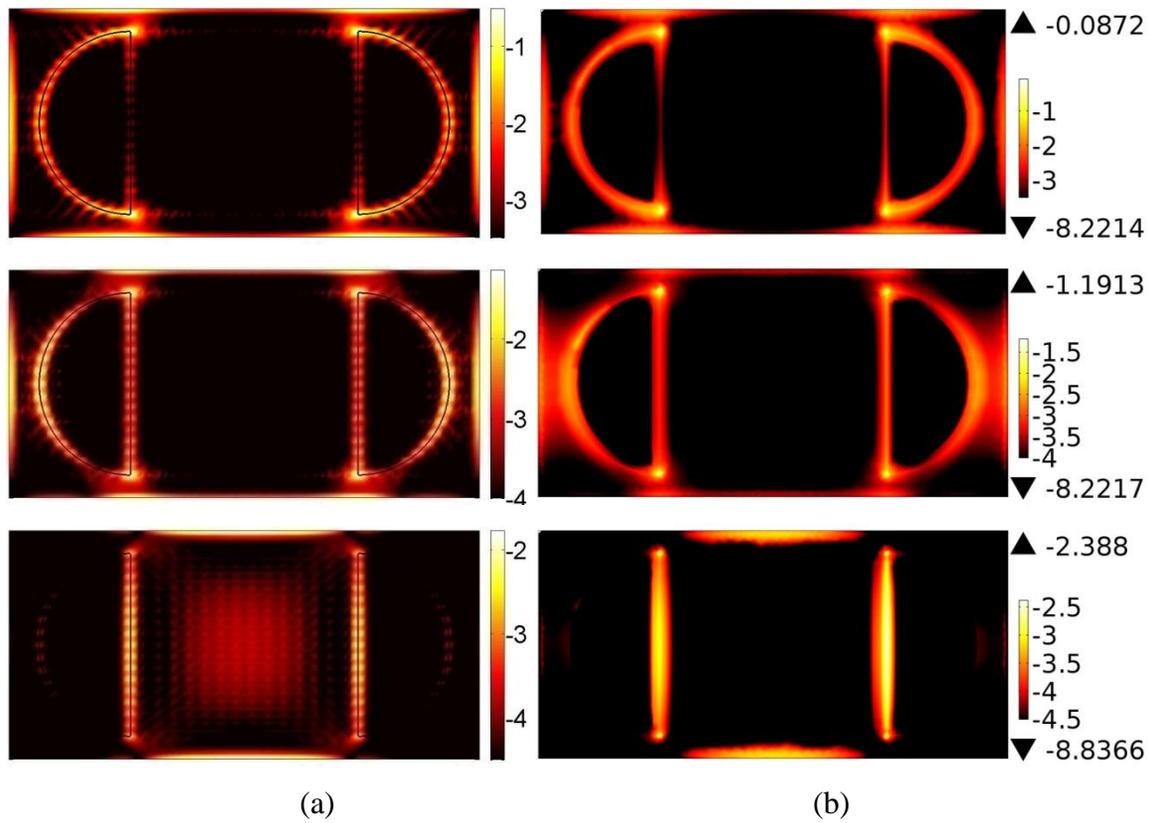

(a)            (b)

FIG. 6

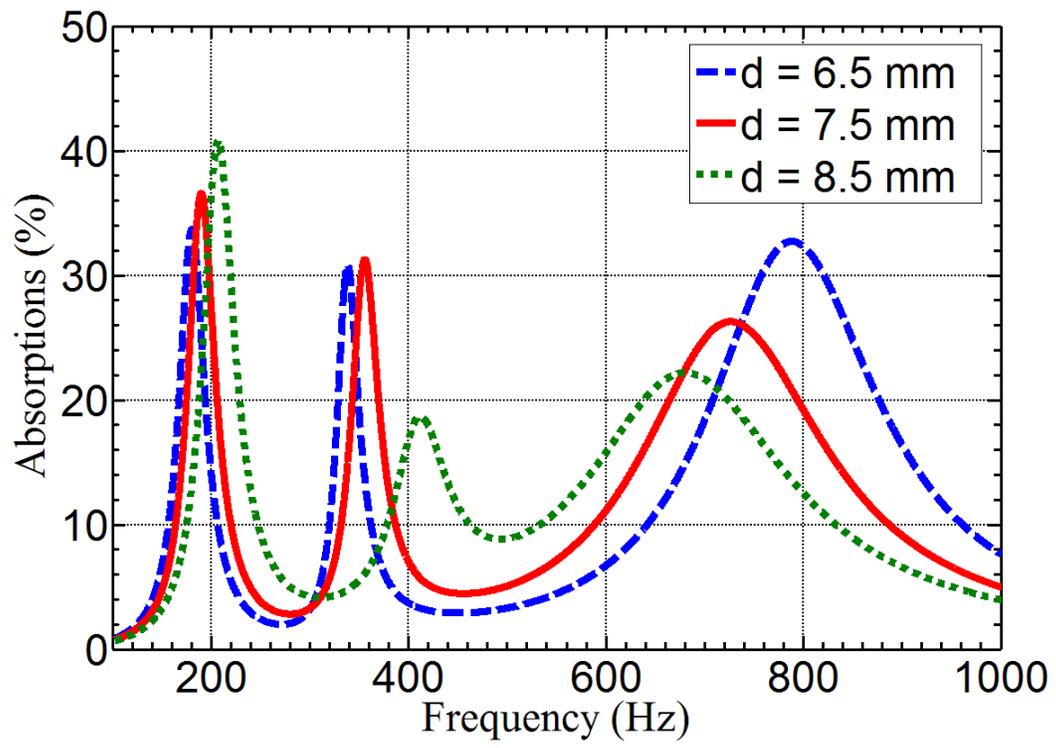

FIG. 7

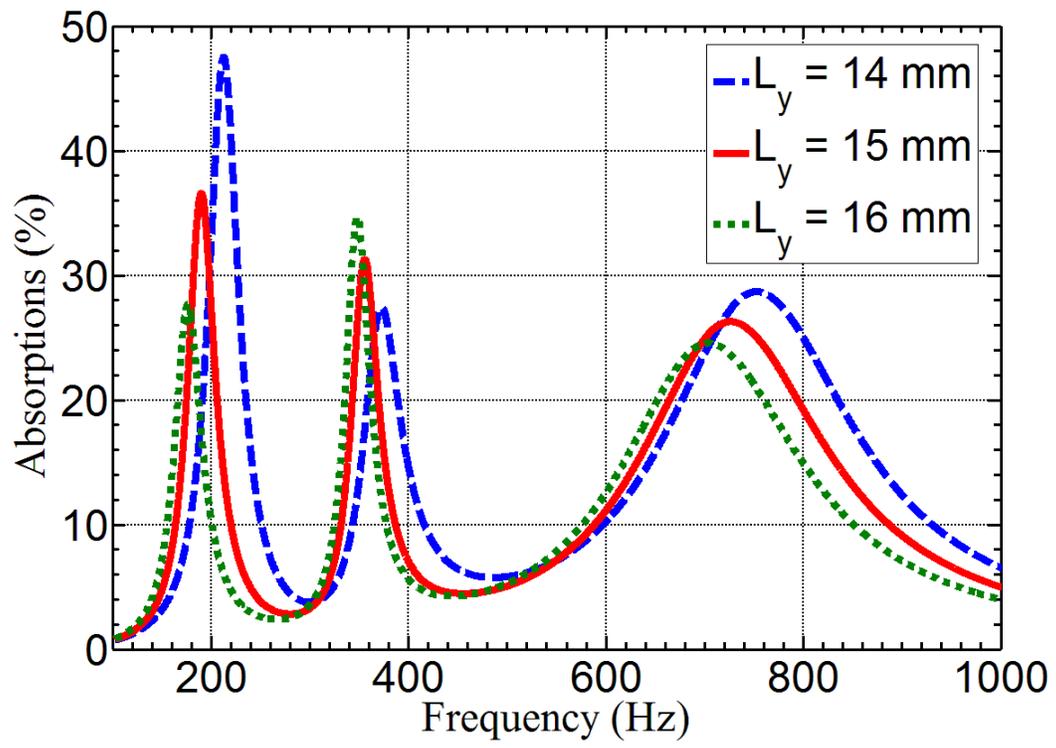

FIG. 8

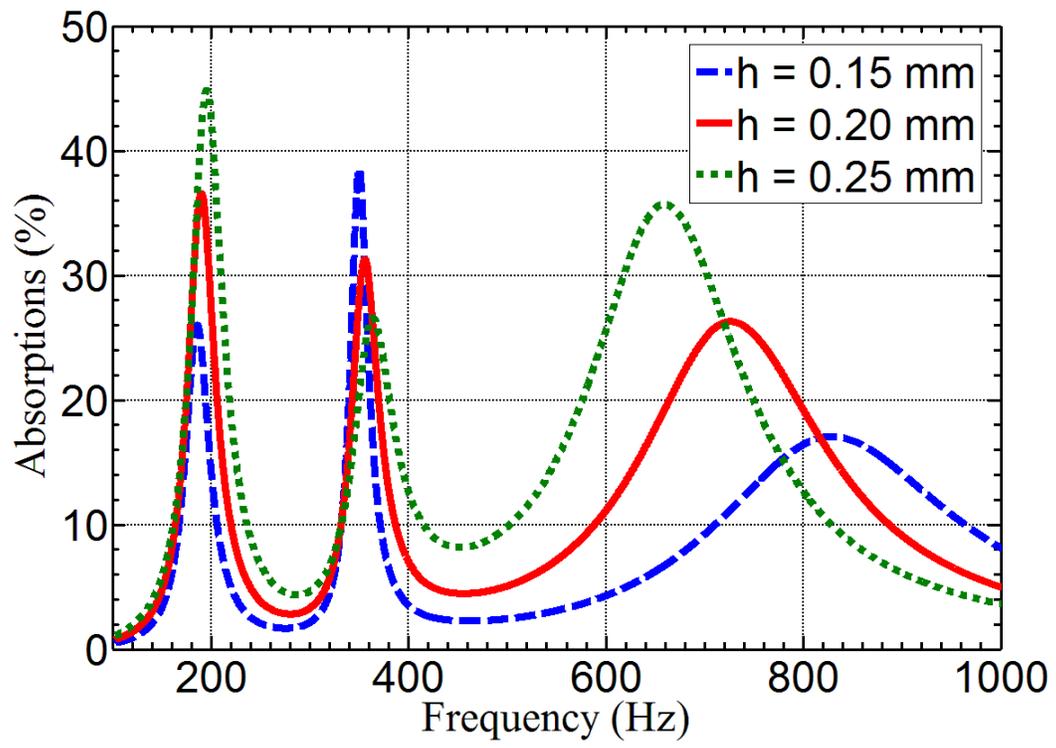

FIG. 9

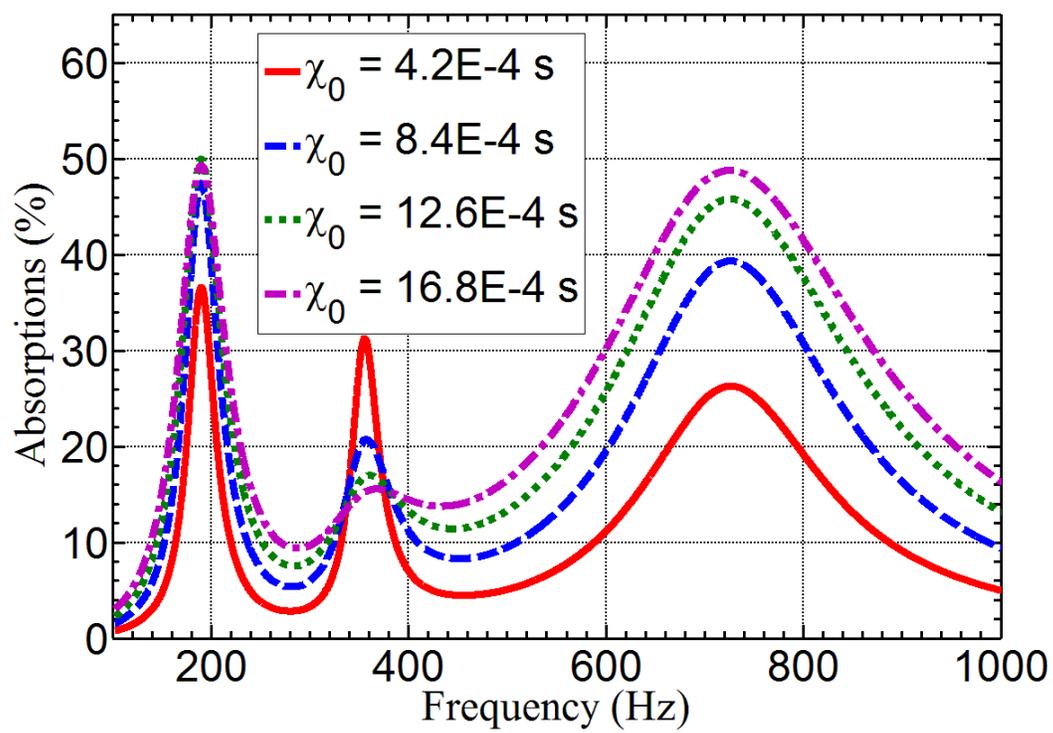

FIG. 10
29

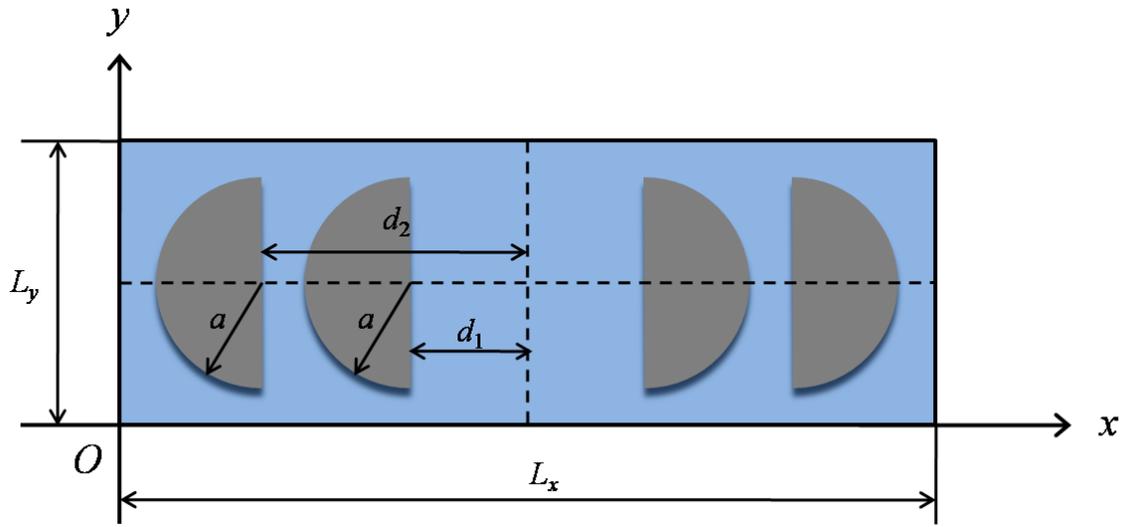

FIG. 11

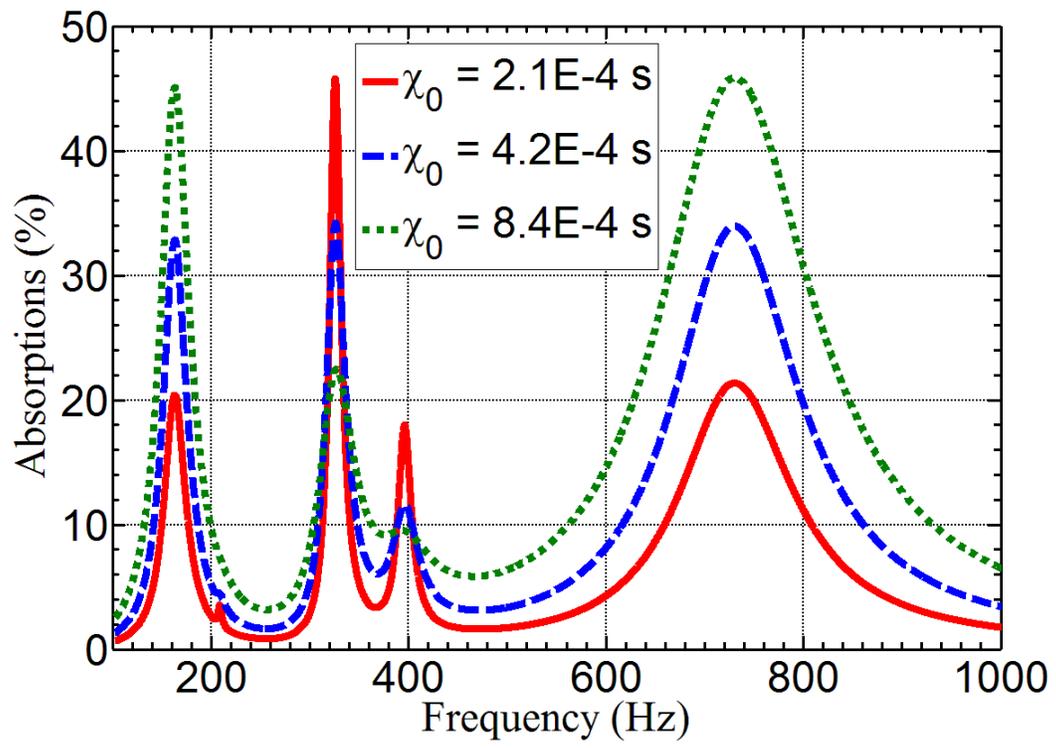

FIG. 12

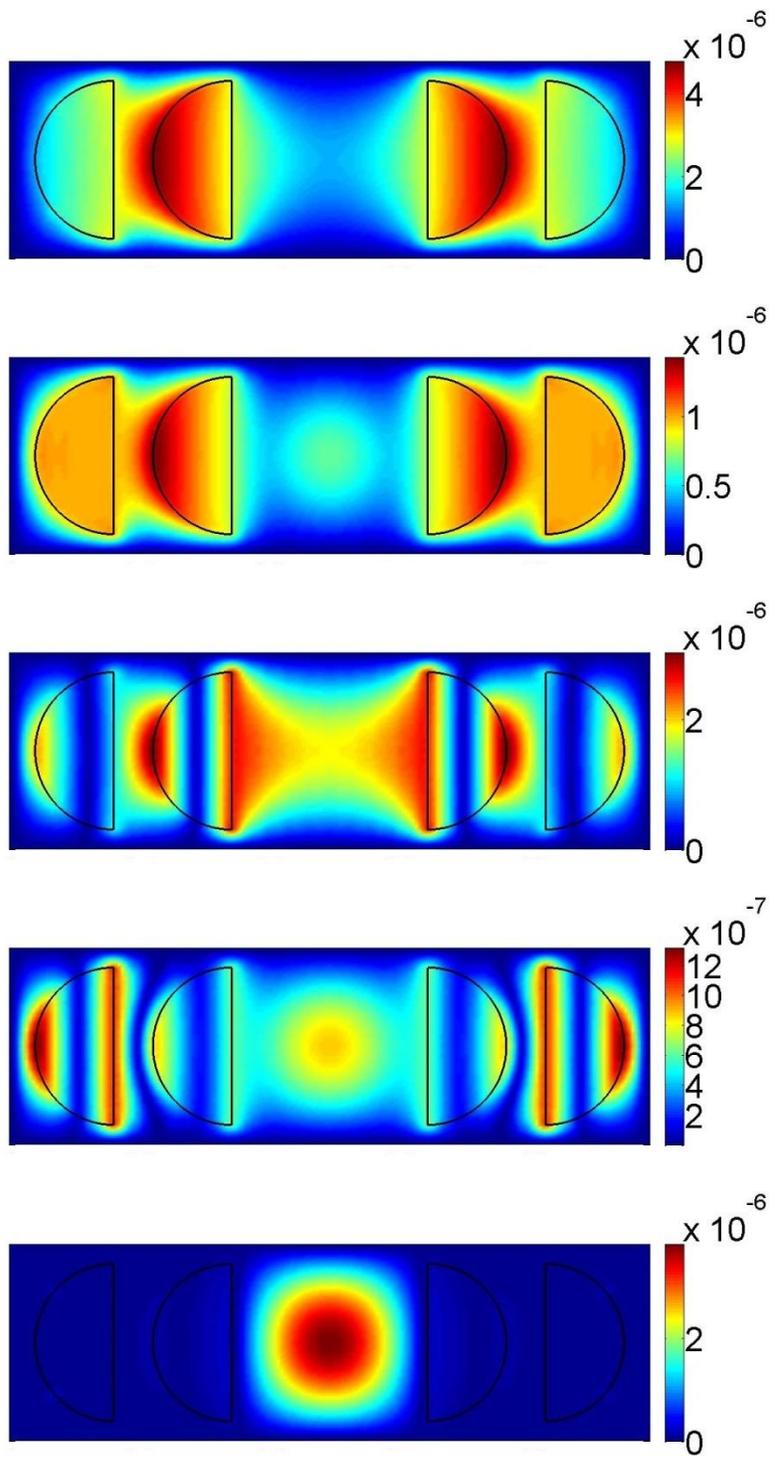

FIG. 13